\documentclass[final,3p,times,twocolumn]{elsarticle}

\usepackage{graphicx}
\usepackage{amssymb}
\usepackage{amsmath}
\usepackage{url}

\usepackage{xcolor}

\biboptions{sort&compress}

\journal{Nuclear Instruments and Methods, Section B}

\begin{document}

\begin{frontmatter}


\title{Uncertainty minimization in electronic stopping cross-section\\measurements using the backscattering method}



\author[usp]{Victor Pires}
\author[usp]{Arilson Silva}
\author[usp]{Cleber L. Rodrigues}
\author[usp]{Nemitala Added}
\author[usp]{Manfredo H. Tabacniks}
\author[usp]{Tiago F. Silva\fnref{myfootnote1}}
\fntext[myfootnote1]{Corresponding author. e-mail: tfsilva@usp.br}
\author[ipen]{\\Flávio Matias}
\author[ipen]{Helio Yoriyaz}
\author[ipen]{Julian Shorto}

\address[usp]{Instituto de Física da Universidade de São Paulo. Rua do Matão, 1371 - Cidade Universitária - São Paulo - Brazil.}
\address[ipen]{Instituto de Pesquisas Energéticas e Nucleares. Avenida Professor Lineu Prestes, 2242 - Cidade Universitária - São Paulo - Brazil.}

\begin{abstract}
Accurate determination of electronic stopping cross sections is critical for ion beam analysis and related applications. While transmission methods are well established, backscattering approaches remain less explored from a metrological perspective, often lacking a systematic treatment of uncertainties. In this work, we present a quantitative framework to optimize experimental geometry in backscattering-based stopping measurements, explicitly accounting for both statistical and systematic errors. Applying the method to helium ions in gold thin films, we identify angular conditions that balance precision and accuracy, achieving total uncertainties below 3\% over a wide energy range. The results, benchmarked against SRIM and ICRU-49, demonstrate that our approach improves the reliability of RBS-derived stopping data and strengthens their use for reference purposes and model validation.
\end{abstract}


\end{frontmatter}


\section{Introduction}
\label{intro}

    The accurate determination of electronic stopping cross sections is essential for a broad range of applications in nuclear and materials science, including ion beam analysis (IBA), ion implantation, radiation damage studies, and detector development. The stopping power represents a fundamental quantity in the description of energy loss processes, and its precise knowledge is required both for modeling and for experimental quantification~\cite{jeynes_rbs_2017,colaux_certified_2015}.  Although theoretical understanding of the physical processes involved in energy loss phenomena is advancing \cite{Alfredo,Ahsan,Roth, de_vera_electronic_2023, matias_deeper-band_2024, matias_efficient_2024}, experimental works focusing on traceable experimental measurements become crucial to benchmark models~\cite{moro_traceable_2016}. Accurate experimental data also help to improve semiempirical approaches to produce tables that are used in simulations targeting a wide range of applications.
    
    Historically, several experimental approaches have been developed to measure stopping cross sections, the most established being the transmission method and the Rutherford backscattering spectrometry (RBS)-based method~\cite{bauer_howto,mertens_how_1987}. While the transmission method is often regarded as more direct, it is not exempt from systematic uncertainties, including foil non-uniformity, energy calibration issues, and beam stability~\cite{moro_traceable_2016,mertens_proton_1986}. Moreover, the perception that it offers intrinsically higher accuracy has contributed to an imbalance in the literature regarding the metrological scrutiny of alternative methods.
    
    In particular, despite its wide use in IBA, the backscattering method remains relatively underexplored from a metrological standpoint. Although early works by Bauer~\cite{bauer_howto} and Mertens~\cite{mertens_proton_1986} have laid important foundations for the method, and some efforts have been made to evaluate the accuracy of RBS-derived stopping data~\cite{bergsmann_how_2001, paul_does_2004, pascual-izarra_experimental_2005}, recent literature still lacks systematic frameworks to quantify and minimize both statistical and systematic uncertainties in these measurements. As a consequence, experimental stopping cross-section data obtained via RBS often omit a complete uncertainty budget, which compromises their traceability and utility as reference data~\cite{colaux_certified_2015}.

    There are different methodologies and strategies to determine experimentally the stopping cross-section in backscattering experiments. Some are based on relative measurements to tables of other better-known elements~\cite{andersen_improved_1978, meersschaut_electronic_2024}, some are constrained by a fitting formula~\cite{eppacher_determination_1988, pascual-izarra_experimental_2005}, and others need some thickness markers~\cite{leblanc_measurement_1994}.
    
    This work aims to address this gap by providing a detailed optimization strategy for stopping cross section measurements using the backscattering method. We present a methodology that identifies optimal geometrical configurations to minimize total uncertainty, incorporating both random and systematic components. Our approach builds upon earlier formulations~\cite{bauer_howto,bergsmann_how_2001}, but extends them with explicit error propagation, correlation analysis, and benchmarking against reference models such as SRIM and ICRU tables. The methodology is applied to helium ions backscattered from thin gold films, for which precise reference data are available.
    
    By offering a reproducible and uncertainty-aware approach to stopping power determination via backscattering, we contribute to improving the reliability of this technique and enabling its broader adoption in precision applications.

\section{Methods}

    \subsection{The back-scattering method}
    \label{sec:backscattering}
    
        There are several approaches to measuring stopping cross-section using backscattering data \cite{bauer_howto}. In this work, we adopt the method based on determining the peak width in thin-film measurements, as described in \cite{chu_chapter_1978}.
        Fig. \ref{fig:spectra} shows a typical spectrum of a pure metallic film deposited on a light substrate. The peak is formed by all scattered ions from the surface down to the film-substrate interface. Its width ($\Delta E$) is the difference between the energy of ions scattered by the film atoms at the surface ($K \cdot E_0$, with $K$ being the kinematic factor and $E_0$ the incidence beam energy) and the exit energy of ions re-emerging after being scattered by film atoms at the interface ($E_{out}$). Assuming a homogeneous stopping in the test film along both, the entry and exit paths, the energy difference can be written as: 

        \begin{figure}[htb!] 
            \centering
            \includegraphics[width=1.0\linewidth]{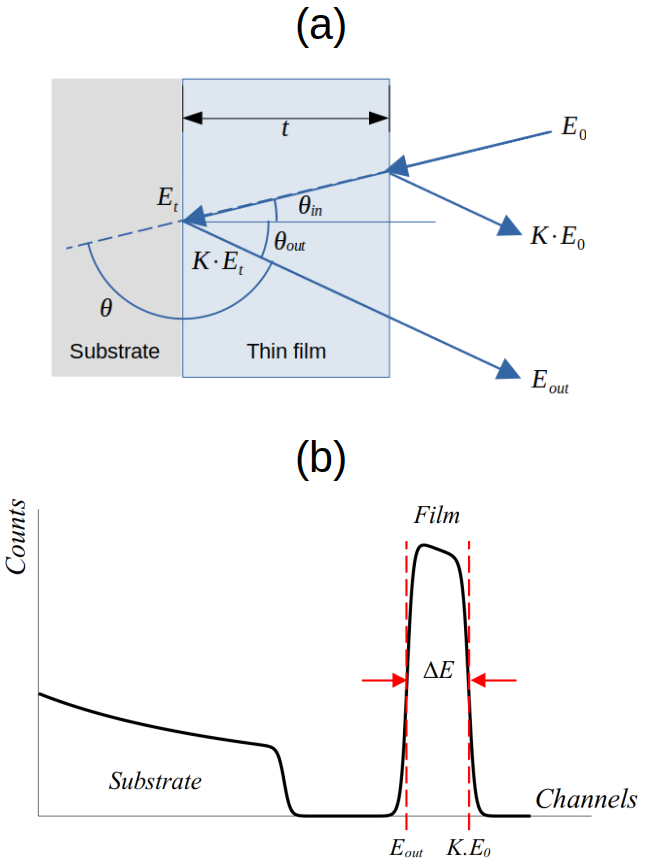} 
            \caption{Illustrative spectrum for the RBS technique.} 
            \label{fig:spectra} 
        \end{figure} 

        The peak width can be determined by adding the energy loss during the path of the ion into the film and the energy loss during its exit from a thin film of thickness $t$. Assuming homogeneous stopping along both the entry and exit paths, this expression can be written as:
        

        \begin{equation}
            \Delta E = \frac{K \cdot t}{\cos(\theta_{in})} [\epsilon]_{in} + \frac{t}{\cos(\theta_{out})} [\epsilon]_{out} 
        \end{equation}
        \\
        with:
        \begin{equation}
            \epsilon_j = \left. \frac{1}{N} \frac{dE}{dx} \right \vert_j
        \end{equation}
        \\
        being the stopping cross section in the $j$ path, either the entry (index $in$) or the exit (index $out$) of the film. $\theta_{in}$ and $\theta_{out}$ represent the angles between the ion's entry and exit paths and the surface normal, respectively. $N$ is the material atomic density.

        The fundamental principle of this methodology is that by measuring the peak width in two different geometrical configurations (i.e. two pairs of values of $\theta_{in}$ and $\theta_{out}$) for the same incident energy, one can obtain a system of equations. The solutions for $[\epsilon_{in}]$ and $[\epsilon_{out}]$ yield the stopping cross section at two different mean energies. In matrix formalism, this can be expressed as: 
        
        \begin{equation}
        \label{eq:system}
             \begin{pmatrix}
            \Delta E^A \\ \\ \Delta E^B \\
            \end{pmatrix} = \begin{pmatrix}
            \frac{K \cdot t}{\cos(\theta_{in}^A)} & \frac{t}{\cos(\theta_{out}^A)} \\
            \\ \frac{K \cdot t}{\cos(\theta_{in}^B)} & \frac{t}{\cos(\theta_{out}^B)} \\
            \end{pmatrix} \begin{pmatrix}
            [\epsilon]_{in} \\ \\ [\epsilon]_{out} \ \\
            \end{pmatrix} 
        \end{equation}
        \\
        where index $A$ and $B$ denote the two different measurements. To better constrain the problem, we impose the conditions $\theta_{in}^A = \theta_{out}^B = \Theta_1$ and $\theta_{in}^B = \theta_{out}^A = \Theta_2$. This assumption is necessary to reduce the number of variables, thereby enhancing result visualization and simplifying the experimental procedure. 
        
        Two things are important to note: the first is that if $\Theta_1 = \Theta_2$ the equation system of Eq. \ref{eq:system} becomes undetermined; and the second is that the factor $K$ implicitly depends on the scattering angle, which remains fixed throughout the analysis. As a consequence, the experimental procedure involves tilting the sample in such a way that, in the first measurement, the incidence angle is $\Theta_1$ and the exit angle is $\Theta_2$, while in the second measurement, the angles are reversed, with the incidence angle set to $\Theta_2$ and the exit angle to $\Theta_1$.

        
        There are different approximations to determine the mean energy at the entry and exit paths of the film \cite{chu_chapter_1978}. Here, we adopt the following approximations:

        \begin{equation}
        \label{eq:mean_energies}
            \begin{split}
            \bar{E}_{in} &= E_0 - \frac{1}{4} \Delta \bar{E}
            \\
            \bar{E}_{out} &= K \cdot E_0 - \left( \frac{2 \cdot K + 1}{4} \right) \Delta \bar{E}
            \end{split}
        \end{equation}
        \\
        with $\Delta \bar{E} = ( \Delta E^A + \Delta E^B) /2 $. This is an alternative definition based on the symmetrical mean energy approximation proposed in \cite{chu_chapter_1978}, which is more convenient for an experimental approach.
        
        Both the assumption of homogeneous energy loss and the approximations used to calculate the mean energy along the ion's entry and exit paths in the film introduce systematic errors. These assumptions do not account for the non-linearity of the stopping cross-section curve. The impact of these approximations on the final result will be evaluated in Section \ref{sec:systematics}.
    
    \subsection{Accounting for random uncertainties} 
    \label{sec:random}

        The random uncertainty in this work arises primarily from the counting statistics of the backscattering spectra, which determine the precision with which the edges of the thin-film peak can be located. In practice, this translates into an uncertainty associated with the discretization of the spectrum into channels, where a uniform probability distribution is assumed within each channel when evaluating the peak width ($\Delta E$). 

        From Eq. \ref{eq:system}, we can propagate the uncertainties in the determination of $\Delta E_j$ to the values of $[\epsilon_j]$ using the definition of a covariance matrix \cite{helene_useful_2016}:
        
        \begin{equation}
        \label{eq:variace}
            V_{\tilde{A}} = \left( X^T V^{-1} X\right)^{-1}
        \end{equation}
        \\
        with:
        
       \begin{equation}
            X = \begin{pmatrix}
            \frac{K \cdot t}{\cos( \Theta_1)} & \frac{t}{\cos( \Theta_2)} \\
            \\ \frac{K \cdot t}{\cos( \Theta_2)} & \frac{t}{\cos( \Theta_1)} \\
            \end{pmatrix} \ \ \ \ \ 
        \end{equation}    

        \begin{equation}
            V = \begin{pmatrix}
            {\sigma_{A}}^2 & 0\\
            \\
            0 & {\sigma_{B}}^2 \\
            \end{pmatrix}
        \end{equation}
        
        Let $\sigma_A$ and $\sigma_B$ be the uncertainties associated with the measurements of $\Delta E_A$ and $\Delta E_B$, respectively. In this work, we assumed $\sigma_A = \sigma_B = G/\sqrt{6}$, where $G$ is the gain calibration of the spectra (see \ref{app:uncert}). This provides a conservative estimate of the precision at which the peak width can be determined from histogram data.

        The adjusted values of the stopping cross-section of a pair of measurements are then given by Eq. \ref{eq:adjust}.

        \begin{equation}
        \label{eq:adjust}
             \begin{pmatrix}
            [\epsilon]_{in} \\ \\ [\epsilon]_{out} \ \\
            \end{pmatrix} = \left( X^T V^{-1} X\right)^{-1} \cdot X^T \cdot V^{-1} \cdot \begin{pmatrix}
            \Delta E^A \\ \\ \Delta E^B \\
            \end{pmatrix}
        \end{equation}

        Note that arranging the matrix $X$ in this format implies alternating $\Theta_1$ and $\Theta_2$ as angles of the entry and exit paths in two consecutive measurements at the same incidence beam energy.
        
        Eq. \ref{eq:variace} results in a matrix whose diagonal holds the uncertainty of our system:
        
        \begin{equation}
        \sigma_{[\epsilon]_{in}} = \frac{G}{\sqrt{6} K t} \cdot \frac{ \sqrt{ \sec ^2 (\Theta_1) +  \sec ^2 (\Theta_2)} } { | \sec^2 (\Theta_1) - \sec^2 (\Theta_2) | }
        \label{eq:var_in}
        \end{equation}

        \begin{equation}
        \sigma_{[\epsilon]_{out}} =  \frac{G}{\sqrt{6} t } \cdot \frac{ \sqrt{ \sec^2 (\Theta_1) +  \sec^2 (\Theta_2)} }{  | \sec^2 (\Theta_1) - \sec^2 (\Theta_2) | }
        \label{eq:var_out}
        \end{equation}
        \\
        and the outer diagonal results in the covariance between the $[\epsilon]_{in}$ and $[\epsilon]_{out}$ values, which is used to calculate the correlation factor $\textbf{r}$ as: 

        \begin{equation}
            \textbf{r} = \frac{\textbf{cov}([\epsilon]_{in}, [\epsilon]_{out}) }{  \sigma_{[\epsilon]_{in}} \cdot \sigma_{[\epsilon]_{out}} } = -{} \frac{ 2 \sec(\Theta_1)\sec(\Theta_2)}{ \sec^2 (\Theta_1) + \sec^2 (\Theta_2)}
        \end{equation}

        Note that the correlation factor is always negative, whatever the combination of $\Theta_1$ and $\Theta_2$.

    \subsection{Accounting for systematic error} 
    \label{sec:systematics}

        In this work, the systematic error refers to the deviation introduced by approximations used in the analytical method, such as assuming a constant stopping cross section along the ion path and estimating the mean energy using simplified expressions. These deviations are quantified by comparing the analytical results with reference values obtained from numerical integration of the SRIM stopping cross-section data.

        To estimate the systematic uncertainties, we tested Eqs. \ref{eq:mean_energies} and \ref{eq:adjust} against the SRIM curve. The peak width, $\Delta E$, has been computed directly using the integral (as in Eq. \ref{eq:integral}) form and SRIM data, while the methodology described in Section \ref{sec:backscattering} is applied to determine $[\epsilon]$ using Eq. \ref{eq:adjust}. These values are then compared to the SRIM values themselves interpolated at the mean energies given by Eq. \ref{eq:mean_energies}. The differences between the interpolated and the output of the matrix calculation are considered as systematic errors. This accounts for systematic errors inserted either by the surface approximation or by the calculation of the mean energy in the entry and exit paths.

        \begin{equation}
        \begin{split}
            E_1 &= E_0 - \int_{0}^{t/\cos(\theta_{in})} { \left[ \frac{1}{N} \frac{dE}{dx} (E(x),x) \right]_{E(0)=E_0} dx } 
            \\
            E_{out} &= K \cdot E_1 - \int_{0}^{t/\cos(\theta_{out})} { \left[ \frac{1}{N} \frac{dE}{dx} (E(x),x)  \right]_{E(0)=K \cdot E_1} dx }
            \\
            \Delta E &= K \cdot E_0 - E_{out}
        \end{split}
        \label{eq:integral}
        \end{equation}

        This approach is justified because SRIM data provide a reliable approximation of the stopping values in light of the current knowledge status of this process. A new measurement aims at improving this knowledge status by restricting the uncertainties. Since our objective is to minimize uncertainties, this method helps identify geometries where our assumptions remain valid.
         
        \begin{equation}        
            s({[\epsilon]_j}) = [\epsilon]_j - \textbf{SRIM}(\bar{E}_j)
        \end{equation}
        \\
        with $s({[\epsilon]_j})$ being the systematic error and $\textbf{SRIM}(\bar{E_{j}})$ being the SRIM interpolation at the energy $\bar{E_{j}}$. $j$ is an index that denotes $in$ or $out$.

    \subsection{Other uncertainty sources}
    
        In addition to the statistical and systematic errors explicitly addressed in this work, several other factors contribute to the overall uncertainty in stopping cross-section measurements. These include uncertainties related to the angular alignment of the beam and detector, variations in the material density, energy calibration, detector resolution, and the intrinsic energy straggling of the ions.

        Energy-loss straggling also contributes to the broadening of the backscattering peak. Although not explicitly included in our analytical propagation, its effect largely follows the same trend as the systematic deviations discussed in Section 2.3, i.e., it increases with the effective path length of ions inside the film. Since the aim of the present work is not to provide absolute uncertainty values but rather a metric to optimize measurement geometries, minimizing the systematic error in practice also minimizes the impact of straggling. In this sense, the optimization strategy remains valid even without an explicit quantitative treatment of straggling.
        
        While a comprehensive treatment of all these effects is beyond the scope of this article, we focus here on developing a strategy for optimizing the geometric configuration of the measurement setup. By propagating selected, dominant sources of uncertainty through our analytical framework, namely, the uncertainty in the energy width ($\Delta E$), we identify angular combinations that minimize the propagated error. A key feature of this approach is the explicit treatment of the correlation between entry and exit path measurements, a factor that is often neglected in conventional stopping cross-section evaluations, yet plays a critical role in minimizing combined uncertainties.
        
        An additional and important source of uncertainty arises from the characterization of the film thickness. While this quantity must be accurately determined using independent experimental techniques (e.g., profilometry or Rutherford Backscattering Spectrometry), its uncertainty should be quadratically summed to the final error obtained by our method. It is important to highlight, however, that the absolute value of the film thickness not only contributes as a multiplicative factor to the stopping cross section determination, but also directly affects the propagation of both statistical and systematic errors, as discussed in Section \ref{sec:random} and \ref{sec:systematics}. This effect is particularly significant for the systematic component, due to its dependence on the energy loss along the ion trajectory.

    \subsection{Finding the best geometry}
    
        With the definitions of random and systematic uncertainties for the case of determination of $[\epsilon]$, the total uncertainty can be defined as:
        
        \begin{equation}
            \sigma_{total} = \sqrt{ s({[\epsilon]_j})^2 +\sigma_{[\epsilon]_j}^2}
        \end{equation}
        
        Note that the common factor in the statistical uncertainty of $[\epsilon]_{in}$ and $[\epsilon]_{out}$, as in Eq. \ref{eq:var_in} and \ref{eq:var_out}, can be written as the function $f(\Theta_1,\Theta_2)$:
    
        \begin{equation}
            f(\Theta_1, \Theta_2) = \frac{ \sqrt{ \sec ^2 (\Theta_1) +  \sec ^2 (\Theta_2)} } { | \sec^2 (\Theta_1) - \sec^2 (\Theta_2) | }
        \end{equation}
        \\
        which diverges when $\Theta_1 \approx \Theta_2$. In special, when $\Theta_1 = \Theta_2$ the system in Eq. \ref{eq:system} becomes undetermined.  $f(\Theta_1,\Theta_2)$ reaches a minimum at $\Theta_1 = 0$ and $\Theta_2 = \pi/2$ or $\Theta_1 = \pi/2$ and $\Theta_2 = 0$. However, as $\Theta_1$ or $\Theta_2$ approaches $\pi/2$, the length of the corresponding ion's path at entry or exit increases, leading to greater systematic errors.

        Therefore, an optimal choice of $\Theta_1$ and $\Theta_2$ must balance reducing random uncertainties while also minimizing systematic errors. Since systematic uncertainty cannot be evaluated analytically, this procedure must be performed numerically.
    
\section{Results and discussion}

    To illustrate the application of the method, we calculated both the random and systematic uncertainties, along with the correlation factor, to optimize an experiment aiming at measuring the electronic stopping cross-section of gold for helium ions. The adopted film thickness in these calculations was 45 nm ($\approx266 \times 10^{15}$ at./cm$^2$). 

    Observing Eqs.~\ref{eq:system} and~\ref{eq:integral}, both the random and systematic uncertainties depend on the film thickness $t$. In this context, the validity of the method relies on representative measurements of $\Delta E$. This imposes an upper limit on the incident beam energy, as it becomes constrained by the achievable depth resolution—i.e., when $\Delta E$ approaches the detector resolution, the depth information is no longer reliable. Assuming a detector resolution of 12~keV (FWHM), depth resolution calculations using ResolNRA~\cite{mayer_resolnra_2008} indicate that a 45~nm film is above this limitation within our energy range of 500-5100~keV.

    \begin{figure}
        \centering
        \includegraphics[width=1\linewidth]{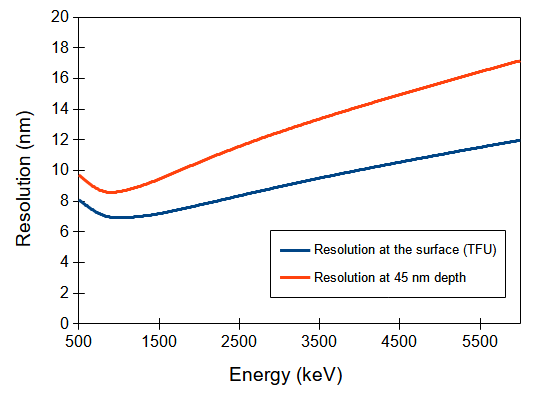}
        \caption{Depth-resolution calculated using ResolNRA demonstrating the 45 nm thick film is suitable for stopping cross-section measurements in the energy range of this study.}
        \label{fig:enter-label}
    \end{figure}
    
\begin{figure*}[h!] 
        \centering
        \includegraphics[width=1.0\linewidth]{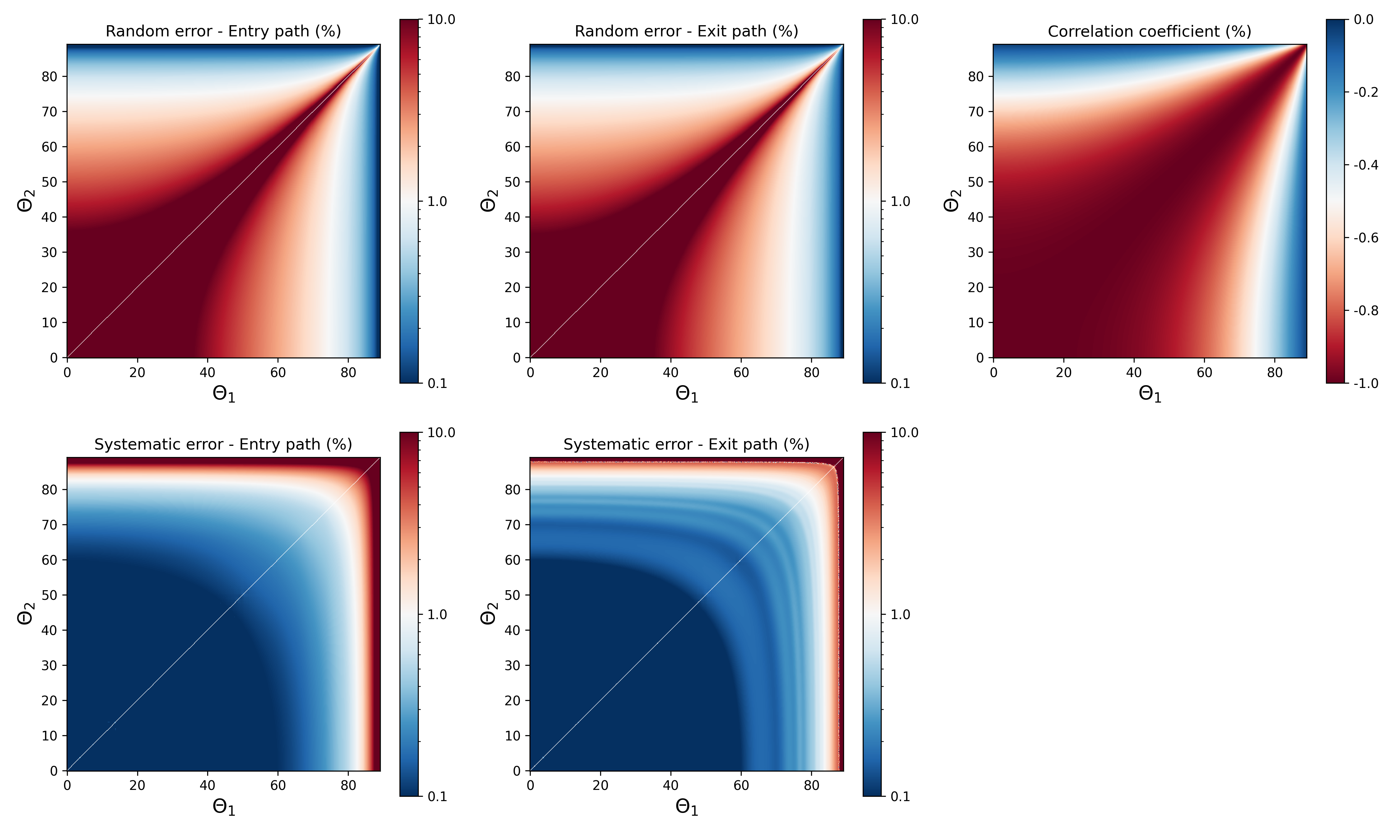} 
        \caption{Panel showing different contributions to the final uncertainty calculated assuming the measurement of helium stopping cross-section in a $3\times10^{17}$ at./cm$^2$ (45 nm) thick gold film. The incident alpha beam energy is 1000 keV and the detector was placed at 120$^\circ$ scattering angle. Top line shows the relative random uncertainty calculated for the way-in (left) and way-out (center) the film, and the correlation factor of these both values (right). The bottom line shows systematic uncertainties for the way-in (left) and way-out (center) the film. Color scale in uncertainty plots are in logarithm scale and saturated at 10\% for improved visualization. The color scale in the correlation factor map in linear and restricted between -1 and 0 since it is always negative.} 
        \label{fig:panel} 
    \end{figure*}

    The results for each contribution to the total uncertainty are presented in the panel of Fig. \ref{fig:panel}.
    
    As mentioned above, the two types of uncertainty exhibit opposite behavior, and the correlation factor presents only negative values (overestimating in the stopping cross section during the ion's entry into the film leads to underestimation during its exit, and vice versa). The correlation factor weakens if $\Theta_1 > 60^\circ$ and $\Theta_2 < 20^\circ$. Uncorrelated measurements are only achievable at specific angle combinations that lead to high systematic errors and are therefore not recommended.

\begin{figure*}[htb] 
        \centering
        \includegraphics[width=0.8\linewidth]{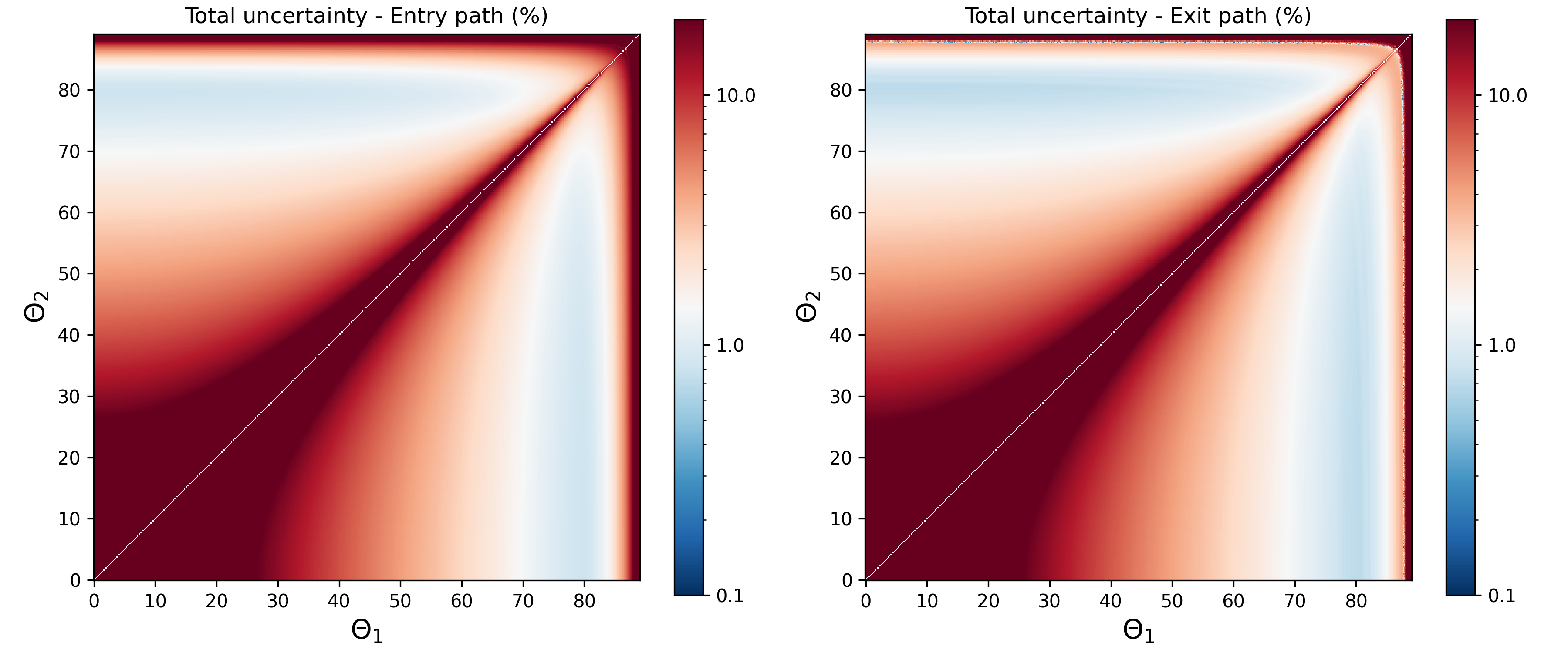} 
        \caption{Heat maps showing calculated total uncertainty assuming the measurement of helium stopping cross-section in a $3\times10^{17}$ at./cm$^2$ (45 nm) thick gold film. The incident energy is 1000 keV and detector placed at 120$^\circ$ scattering angle. Total uncertainty in the way-in (left) and way-out (right). Color scale in uncertainty plots are in logarithm scale and saturated at 10\% for improved visualization.} 
        \label{fig:final_1000} 
    \end{figure*}

\begin{figure*}[htb] 
        \centering
        \includegraphics[width=0.8\linewidth]{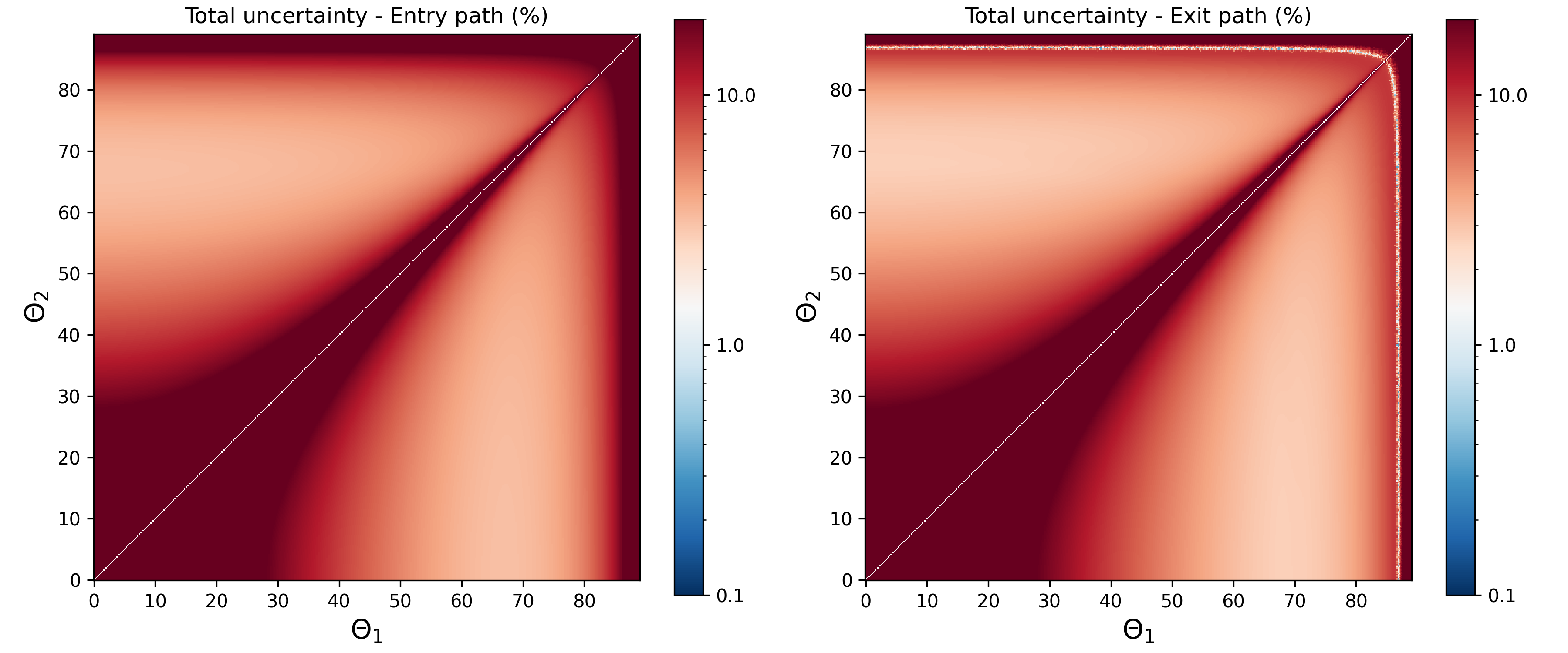} 
        \caption{Heat maps showing calculated total uncertainty assuming the measurement of helium stopping cross-section in a $3\times10^{17}$ at./cm$^2$ (45 nm) thick gold film. The incident energy is 500 keV and detector placed at 120$^\circ$ scattering angle. Total uncertainty in the way-in (left) and way-out (right). Color scale in uncertainty plots are in logarithm scale and saturated at 10\% for improved visualization.} 
        \label{fig:final_500} 
    \end{figure*}

\begin{figure*}[htb] 
        \centering
        \includegraphics[width=0.8\linewidth]{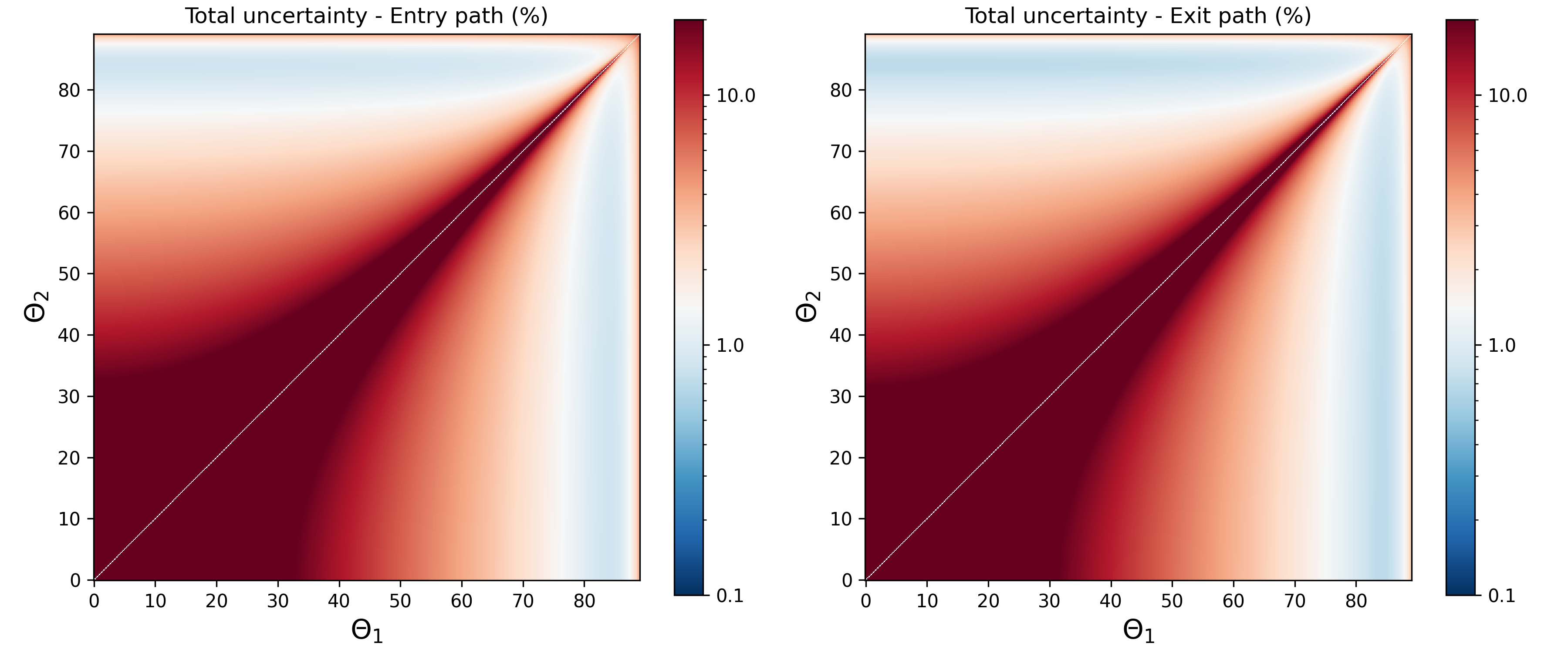} 
        \caption{Heat maps showing calculated total uncertainty assuming the measurement of helium stopping cross-section in a $3\times10^{17}$ at./cm$^2$ (45 nm) thick gold film. The incident energy is 5000 keV and detector placed at 120$^\circ$ scattering angle. Total uncertainty in the way-in (left) and way-out (right). Color scale in uncertainty plots are in logarithm scale and saturated at 10\% for improved visualization.} 
        \label{fig:final_5000} 
    \end{figure*}

    The combination of random and systematic uncertainties yields the total uncertainties as shown in Fig. \ref{fig:final_1000}. We also calculated the final uncertainties for incident energies of 500 keV and 5000 keV, with the results presented in Figs. \ref{fig:final_500} and \ref{fig:final_5000}, respectively.
    
    From these results, it is clear that, to keep the total uncertainty below 3\% in the overall energy range, $\Theta_1$ should be kept between 50$^\circ$ and 80$^\circ$ and $\Theta_2$ should be below 30$^\circ$. Furthermore, it can be observed that the total uncertainties below 1\% are not feasible in this experiment. If higher accuracy is needed, alternative methods should be explored \cite{moro_traceable_2016}.

    Finally, we performed experiments with incident energies ranging from 500 keV up to 5100 keV. The set of angles was chosen as $\Theta_1 = 0^\circ$ and $\Theta_2 = 60^\circ$. This choice results in a maximum total uncertainty of 3\% for measurements with incident energy of 500 keV and a minimum of 2\% for measurements with incident energy of 1000 keV (at the maximum of the stopping cross-section curve). The correlation factor in this set of angles is approximately 0.7.

\begin{figure*}
        \centering
        \includegraphics[width=0.8\linewidth]{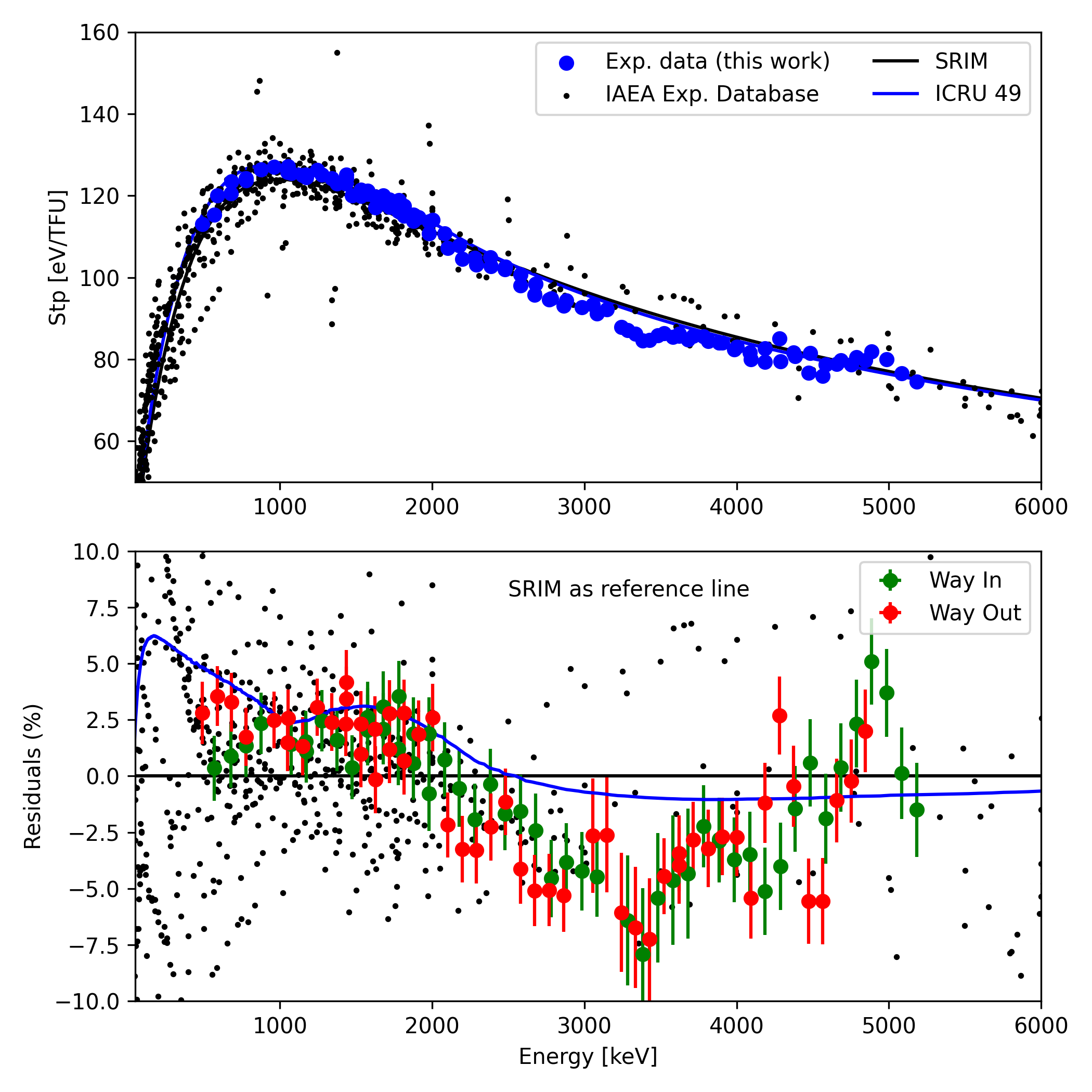}
        \caption{Stopping cross section measured for gold to helium ions (upper plot) and Relative residuals calculated using SRIM as reference (bottom plot). The geometry used was defined using the methodology reported here. 1 TFU = $1 \times 10^{15}$ at./cm$^2$. In the residuals, data was labeled according to the way used in the average, i.e. way in or way out the thin film.}
        \label{fig:stp}
    \end{figure*}

    Fig. \ref{fig:stp} shows the experimental data obtained using the method described in \ref{sec:backscattering} with a combination of angles chosen accordingly to our approach. The results are compared to SRIM and ICRU-49 data. The black points in the plot represent additional experimental data from the IAEA database. In the upper plot, we observe the overall agreement of our data with the semi-empirical models, while in the bottom plot, we display the relative residuals with SRIM as the reference. We provide densely spaced experimental data to better define the statistical limits of the method. At lower energies, it is evident that all experimental data fall within 4\% in residuals amplitude, a value that is very close to the final uncertainty calculated using the method proposed here.

    In the residuals plot in Fig. \ref{fig:stp}, our experimental data were labeled accordingly to determine whether they were obtained using the entry path (green) or the exit path (red). We observe a great consistency of the data determined by both paths.

    The comparison with the models shows reasonable agreement in the limit of uncertainty bars both for SRIM and ICRU-49, with the latter better capturing the fine structure of data distribution in the residuals plot.

\section{Conclusions}

    In this work, we have presented a novel methodology for optimization when measuring stopping cross-section using backscattering data, offering an accurate and reliable approach for determining the stopping cross-section in thin films. Our method, which is based on a systematic treatment of uncertainties, provides a robust framework for minimizing random and systematic errors while optimizing geometries for precise measurements.

    It is worth noting that energy straggling, although not explicitly included, is expected to follow the same angular dependence as the systematic uncertainties, and thus its effect is indirectly minimized by the present optimization strategy.

    The experimental results, which were obtained for helium in gold in a wide range of incident energies, show good agreement with well-established semi-empirical models, such as SRIM and ICRU-49, and highlight the effectiveness of our approach in achieving uncertainties that are close to the theoretical limits. The careful selection of geometries and the use of densely spaced data points were crucial in refining the statistical accuracy of our measurements, particularly at lower energies.

    Although total uncertainties below 1\% were not feasible in this experiment, the method provides a reliable tool to plan stopping cross-section measurements, allowing better restriction of the spread in obtained experimental values.

\section*{Acknowledgments}

    The authors thank the financial support provided by CNPq-INCT-FNA (project number 464898/2014-5). TFS acknowledges the Brazilian funding agency CNPq (project number 406982/2021-0). FM acknowledges CNEN (Project No. 2020.06.IPEN.32). This research used resources from the Laboratory for Materials Analysis with Ion Beams - LAMFI-USP,  of the University of S\~ao Paulo. The authors acknowledge laboratory staff for their assistance during the experiments.

\hfill \break
During the preparation of this work, the authors used ChatGPT by OpenAI to improve language and readability. After using this tool, the authors reviewed and edited the content as needed and assume full responsibility for the content of the publication.
    

\appendix
\section{Random uncertainties in $\Delta E$ determination}
\label{app:uncert}
To determine the thin film peak width at half of its maximum, it is necessary to determine in which channels the half height occurs in the trailing and falling edges. Being conservative, we can assume a uniform probability distribution within one channel, leading to $\sigma_{\texttt{edge}} = 1/\sqrt{12}$. With $\Delta E$ calculated as the difference in channels of the trailing and falling edges times the gain factor $G$, the uncertainty of $\Delta E$ can be calculated as:

\begin{equation}
    \sigma_{\Delta E} = G \cdot \sqrt{ \left( \frac{1}{\sqrt{12}} \right)^2 + \left( \frac{1}{\sqrt{12}} \right)^2 } = \frac{G}{\sqrt{6}}
\end{equation}




\bibliographystyle{model1-num-names}
\bibliography{references.bib}







\end{document}